\documentclass[times,twocolumn,final,10pt,a4paper]{elsarticle}
\pdfoutput=1

\usepackage{latexsym,amsmath,amssymb,amsfonts}

\usepackage{url,color}

\newcommand{\set}[1]{\mathbb #1}

\newcommand{\fxp}[2]{\text{fx}{\ensuremath{#1.#2}}}

\DeclareMathOperator{\argmin}{arg\,min}

\def\setU{\set{U}}
\def\setS{\set{S}}
\def\setC{\set{C}}
\def\setCp{\set{C'}}
\def\setD{\set{D}}

\definecolor{red}{rgb}{1.0,0,0}
\definecolor{green}{rgb}{0,1.0,0}
\definecolor{blue}{rgb}{0,0,1.0}

\journal{Pattern Recognition Letters}

\begin{document}

\begin{frontmatter}

\title{Geometric Near-neighbor Access Tree (GNAT) revisited}

\author{Kimmo Fredriksson}
\address{School of Computing, University of Eastern Finland,
P.O. Box 1627, 70211 Kuopio, Finland}
\ead{kimmo.fredriksson@uef.fi}

\begin{abstract}
Geometric Near-neighbor Access Tree (GNAT) is a metric space indexing method based on 
hierarchical hyperplane 
partitioning of the space. While GNAT is very efficient in proximity searching, 
it has a bad reputation of being 
a memory hog. We show that this is partially based on too coarse analysis, and that the memory
requirements can be lowered while at the same time improving the search efficiency. 
We also show how to make GNAT memory adaptive in a smooth way, and that the hyperplane 
partitioning can be replaced with ball partitioning, which can further improve the search 
performance. We conclude with experimental results showing the new methods can give 
significant performance boost.
\end{abstract}


\end{frontmatter}


\section{Introduction}

Efficient solutions to similarity or proximity search problem have many increasingly important
applications in several areas, most notably in (multi)media and information retrieval.
Besides the usual database centric model some similarity searching 
methods can be seen as nearest neighbor classifiers as well, and have
applications as internal tools in many systems (e.g.\ for lossy video or audio compression,
pattern recognition and clustering, bioinformatics,
machine learning, artificial intelligence, data mining).
Metric space is a pair $(\setU, d)$,
where $\setU$ is an universe of objects, and $d(\cdot, \cdot)$ is a distance
function $d : \setU \times \setU \to \mathbb R^+$.
The distance function is {\em metric}, if it satisfies for all $x, y, z \in {\mathbb U}$
\begin{align*}
d(x,y)  ={} & 0 \text{ iff } x = y \text{ (reflexivity)}, \\
d(x,y)  ={} & d(y,x) \text{ (symmetry)}, \\
d(x,y)  \leq{} & d(x,z) + d(z,y) \text{ (triangular inequality)}.
\end{align*}
In the point of view of the applications, we have some subset $\setS \subseteq \setU$
of objects, $|\setS|=n$, and we are interested in the proximity of the objects towards themselves, 
or towards some query objects. 
The most fundamental type of query is {\em range query}: retrieve all objects in the 
database $\setS$ that are within a certain similarity threshold $r$ to the given query 
object $q$, that is, compute $R(\setS,q)=\{o \in \setS\;|\;d(q,o) \leq r\}$. 
Another common query (which can be solved with suitably adapted range query as well) 
is to retrieve the $k$-nearest neighbors of $q$ in $\setS$. 
A large number of different data structures and query algorithms have been proposed, see 
e.g.\ \cite{CNBYMacmcs01.2,idriven,Sametbook,Zezulabook}.

\subsection{Contributions}

In this paper we take a fresh look on the well-known GNAT \cite{gnat} data structure. 
While it has some attractive properties, it is often dismissed as having too large memory 
requirements (which is partially based on too coarse analysis, as we show). 
We give several techniques to improve the space complexity, make it memory 
adaptive in a way that is arguably more elegant than in the baseline GNAT, 
improving its search performance on the same time. 
We also show that it is possible to replace GNAT's hyperplane partitioning with 
ball partitioning, which gives more flexibility in certain situations and can also further 
improve the performance. 
It is also possible to increase the tree arity while keeping the same memory usage.
Recently GNAT gained new interest also in the form of EGNAT \cite{egnat}, a dynamic and
external memory based variant of GNAT. Many of our techniques can benefit EGNAT as well,
and we discuss some methods that can improve construction and insertion costs in our GNAT 
variant.
We conclude with experimental results that show substantial improvements in space usage 
and query performance.

\section{AESA, GNAT and EGNAT}

We briefly review the algorithms relevant to the present work. Our work is based on GNAT, but
GNAT itself has some connections to AESA (which we will make more explicit in what follows). 
EGNAT is a dynamic external memory variant of GNAT.
We also give a new analysis for GNAT in this Section.

\subsection{AESA}

Approximating Eliminating Search Algorithm (AESA) \cite{aesa} is one of the most well-known and
one of simplest approaches to index a metric space. It is also the best in terms of number of 
distance evaluations needed to answer range or $k$-NN queries. The drawbacks are its quadratic space
requirement and high extra CPU time (the time spent on other work than pure distance evaluations).

The data structure is simply a precomputed matrix of all the $n(n-1)/2$ distances between the $n$ 
objects in $\setS$. The space complexity is therefore $O(n^2)$ and the matrix is computed with
$O(n^2)$ distance computations. This makes the structure highly impractical for large $n$.

The range query algorithm is also simple. First the distance $e=d(q,p)$ between the query $q$ and
randomly selected pivot $p \in \setS$ is evaluated. If $e \leq r$, then $p$ is reported. 
Then each object $u \in \setS \setminus \{p\}$ that
does not satisfy $e-r \leq d(u,p) \leq e+r$ is eliminated, i.e.\ we compute a new set 
$\setS' = \{ u \in \setS \setminus \{p\}~|~e-r \leq d(u,p) \leq e+r\}$. Note that the distances $d(u,p)$ can
be retrieved from the precomputed matrix. However, the elimination
process has to make a linear scan over the set $\setS$, so the cost is the
time for one distance computation plus $O(n)$ extra CPU time. This process is
repeated with a new pivot $p$ taken from the qualifying set $S'$. This selection can be random, or 
e.g.\ the one that minimizes the lower bound distance to $q$ (which can be maintained during the 
search with constant factor overhead). This is repeated until $\setS'$ becomes empty.
By experimental results \cite{aesa} the search algorithm makes only a
constant number distance computations on average, which means $O(n)$
extra CPU time on average. One should note that the $O(1)$ result means that $n$
does not affect the number of distance evaluations, while the ``constant'' has exponential dependence on the dimension of the space and the search radius.

There are many approaches to reduce the space and/or the extra CPU time 
(e.g.\ \cite{laesa,baesa,t-spanners,tlaesa,spaghettis,roaesa}), but these induce more
distance computations or extra CPU time or work only for $k$-NN queries (\cite{roaesa}). 

\subsection{GNAT} \label{sec:gnat}

Geometric Near-neighbor Access Tree (GNAT) \cite{gnat} 
is based on hyperplane partitioning applied recursively to obtain an $m$-ary
tree (where $m$ is a constant / parameter). The tree is built as follows:
\begin{enumerate}
\item
Select $m$ {\em centers} or pivots 
(called split points in \cite{gnat})
$\setC = \{c_1, \ldots, c_m\} \subseteq \setS$.

\item
Associate each object in $\setS \setminus \setC$ with the closest center in $\setC$, obtaining
sets $\setD_{c_i}$.

\item
Compute a {\em distance range table} $R$ for the current node, where $R_{i,j} = [lo,hi]$,
defined as 
\[
R_{i,j} = [\min\{d(c_i, x)\;|\;x \in \set{X} \}, \max\{d(c_i, x)\;|\;x \in \set{X} \}],
\]
where $\set{X} = \setD_{c_j} \cup \{c_j\}$.
\item
Build the children $i \in \{1 \ldots m\}$ recursively using the same method for $D_{c_i}$.
\end{enumerate}
Thus the centers $\setC$ induce a recursive Voronoi partitioning of the space. 
The centers can be selected at random, or using some heuristic \cite{gnat} method. 
It is also possible (and common) to stop the recursion when some predefined number of $b$
objects are left (i.e.\ $|\setD| \leq b$), and simply store the remaining objects as a bucket 
in the leaf.

Notice that if for $R_{i,j}$ the sets $\setD_{c_j}$ are empty, then $lo = hi$ and the range 
table in fact becomes (almost) the same matrix that AESA uses. The difference is that $R$ 
uses more space, as each $lo$ value is a duplicate of the corresponding $hi$ value, 
and AESA uses the symmetry
$R_{i,j} = R_{j,i}$ (in this special case) to store only half of the matrix.

GNAT is often cited to have $O(nm^2)$ space complexity, which is an obvious upper bound,
but much tighter bound of $O(nm)$ can also be derived.
We give more detailed analysis below. We assume that the tree is balanced, i.e.\ has 
depth $\log_m n$. This is hard to guarantee with hyperplane partitioning, but nevertheless 
in practice the tree tends to have $O(\log_m n)$ average depth \cite{egnat}. 

The space complexity can be expressed as\footnote{Or more accurately,
$S(n)=m S((n-m)/m) + m + m^2 = m S(n/m-1)+m+m^2$, but we use the simpler formula which gives 
an upper bound.}  
$S(n) = m S(n/m) + m^2$ and by substituting we get
\begin{align*}
S(n)  ={} & m S(n/m) + m^2 \\
      ={} & m^2 S(n/m^2) + m^3 + m^2 \\ 
      ={} & \ldots \\
      ={} & m^i S(n/m^i) + \sum_{j=1}^i m^{j+1} \\
      ={} & m^i S(n/m^i) + m^2 \sum_{j=1}^i m^{j-1} \\
      ={} & m^i S(n/m^i) + m^2 (m^i-1)/(m-1). 
\end{align*}
By noticing that the recursion stops when $i=\log_m n$ 
we obtain $S(n) = O(nm)$. 
On the other hand, if the tree is extremely unbalanced, say, $m-1$ of the children are empty, 
and all objects go to the remaining branch, the space is $S(n) = S(n-m) + m^2$. The solution 
is simple, again 
$S(n) = O(m^2 n/m) = O(nm)$.

The time complexity to build the tree is 
$T(n) = m T(n/m) + m^2 n/m = m T(n/m) + mn$. It is easy to see that the time spent in each level
of the (recursion) tree is $O(nm)$, and hence the whole process takes $O(nm \log_m n)$ time.
In the worst case (for unbalanced tree) the time becomes $T(n) = T(n-m) + O(nm) = O(n^2)$.

As we will see later, using non-constant arities of the the form $n^\alpha$ in the nodes, for 
some $\alpha \leq 1$, gives much better query performance while using the same space as the 
standard constant arity variant. The price to pay is even costlier preprocessing step in general.

The range search algorithm for a query object $q$ and range $r$ resembles the method that
AESA uses, except that it is recursive in GNAT, starting from the root of the tree:
\begin{enumerate}
\item
Select a pivot $p_i \in \setC$ (each pivot at most once).
\item 
Compute 
$e = d(p, q)$, and if $e \leq r$, then report $p$.
\item
If the query ball does not intersect the range associated with some center, then the center
(and the corresponding subtree) can be pruned, 
i.e.\ if $[e - r, e + r] \cap R_{i,j}$ is empty, we can eliminate $p_j$; or in other words,
if $R_{i,j} = [lo,hi]$, then if 
$e - r > hi$ or $e + r < lo$ we can eliminate $p_j$.
\item 
Repeat the steps 1--3 until all surviving centers have been tried.
\item 
For each survived center $p_i$ search recursively the corresponding subtree $D_{p_i}$.
\end{enumerate}
Again notice that if the subtrees are empty, then the above algorithm ``degenerates'' into AESA.

\subsection{EGNAT}

As presented, GNAT is a static structure once built. EGNAT \cite{egnat} is a dynamic version of 
GNAT that allows insertions (relatively easy) and deletions (needing novel ideas) of objects.
EGNAT is also designed to work efficiently with external memory (disk) setting.

We do not go into the details, except for the search algorithm (ignoring the modification that 
is needed to handle deletions of objects). EGNAT uses the same distance range tables as GNAT, 
but uses otherwise simpler (and less powerful) algorithm.

More precisely, given a query object $q$, the EGNAT search algorithm computes {\em all} the distances
$d(q,c_i)$, reporting $c_i$ if $d(q,c_i) \leq r$, and finding the nearest neighbor of $q$ among the 
centers in $\setC$, i.e.\ it computes $j = \argmin_{1 \leq i \leq m} d(q,c_i)$ and records 
$e = d(c_j,p)$. Then all $m$ centers are scanned sequentially, and the children $i$ is pruned
if $|e-r, e+r|$ does not intersect the range $R_{i,j}$. Note that this method has lower extra CPU time
than GNAT, but in general makes more distance evaluations, and the difference increases when the tree
arity increases. On the other hand, EGNAT uses relatively low arities.

\section{Improved GNAT}

We present several orthogonal improvements to the baseline GNAT. 
All of them can be combined into a single algorithm.

\subsection{Non-constant arities} \label{sec:powa}

Assume now that the node arities are of the form $\sqrt{n_i}$, where $n_i$ is the number of objects
associated on a node at level $i$ of the tree. If the tree is balanced, the space becomes now
\[S(n) = \sqrt{n}\, S(\sqrt{n}) + n.\] 
The solution is $S(n) = O(n \log\log n)$, which is easy to 
see by writing the recurrence as 
\[S(n) = 2^{(\log_2 n) / 2}\, S(2^{(\log_2 n) / 2}) + n.\]
That is,
the tree has depth $\log\log n$ now, and the total space per tree level is $O(n)$. 
The time complexity to build the tree is $T(n) = \sqrt{n} T(\sqrt{n}) + n^{1.5}$, which solves
to $O(n^{1.5} \log\log n)$.
As we see in Sec.~\ref{sec:exp}, this performs much better than using constant arity, where 
$m \approx \log\log n$.
Note that as a side effect this method also gears the tree towards balance, as the arities
are ``automatically'' higher for nodes that have many objects associated with them.

If the node arities are of the form $n^\alpha$, for $0 < \alpha \leq 1$ (where above we considered 
the case $\alpha = \frac{1}{2}$), the space becomes 
\[S(n) = n^\alpha\, S(n^{1-\alpha}) + n^{2\alpha}.\]
For $\alpha \geq \frac{1}{2}$, the total space per tree {\em level} decreases when $\alpha$ 
increases, which gives a (somewhat loose) upper bound 
\[S(n) = O(n^{2\alpha} \log_{1/(1-\alpha)}\log n).\]
Similarly the construction time becomes 
\[T(n) = n^\alpha T(n^{1-\alpha}) + n^{1+\alpha} = O(n^{1+\alpha}\log_{1/(1-\alpha)}\log n).\]
Notice that for $\alpha = 1$ the tree has only one node (the root) and the space becomes $n^2$, and
we have effectively turned GNAT into AESA. 
Indeed, when $\alpha=1$ the GNAT search algorithm becomes AESA, as in this case all the 
child nodes are empty and hence the $lo$ and $hi$ values in the range tables are equal, 
and the range table matrix also becomes symmetric.
Note that the higher arity we use, the smaller clusters
the hyperplane partitioning produces, and therefore the $lo \ldots hi$ ranges in the range
tables become smaller, which in turn enable more pruning, as the probability of intersecting 
with the (range) query ball becomes smaller.

Thus we have a smooth transition from GHT-like 
tree\footnote{GHT (Generalized Hyperplane Tree) \cite{ght} resembles GNAT with $m=2$.} 
to GNAT
to AESA if we adjust $\alpha$ from $1 / \log_2 n_i$ to $1$.
Likewise,
the preprocessing time grows
from $O(n \log n)$ to $O(n^2)$. 

\subsection{Ball partitioning} \label{sec:ballp}

An interesting point is that while GNAT (preprocessing) is based on hyperplane partitioning, 
the range tables and the search algorithm do not make {\em any} assumption of this. Thus the 
algorithm works as is, even if we replace the hyperplane partitioning with ball 
partitioning\footnote{MVPT (Multi-Vantage-Point-Tree) \cite{mvpt} also uses several pivots
per node and ball partitioning, but the resulting structure is otherwise quite different to GNAT.}. 
This gives some 
interesting 
opportunities to tune the data structure, in particular 
it is easy to control the (un)balance of the tree. 

The drawback as compared to hyperplane partitioning is that the ball clusters can intersect, 
which means that even exact search might need to enter more than one branch of the tree 
during the search. On the other hand, in dynamic settings, such as in EGNAT, ball partitioning
gives more flexibility to handle the insertions.

To implement the ball partitioning, we basically need to only change the preprocessing step 2
(see Sec.~\ref{sec:gnat}) for the baseline GNAT, while the search algorithm stays intact.
That is, we replace the step 2 with the following 
(the arities can still be selected as $m = n_i^\alpha$):
\begin{enumerate}
\item[2 (a).]
Select ball capacities $b_1 \ldots b_{m-1}$.
\item[2 (b).]
For each $i \in \{1 \ldots m-1\}$, in ascending order, select a radius $r_i$ so that $c_i$ with
radius $r_i$ covers exactly $b_i$ objects in 
$(\setS \setminus \setC) \setminus \bigcup_{j<i} \setD_{c_j}$; 
put these into $D_{c_i}$.
\item[2 (c).]
Put the remaining objects $(\setS \setminus \setC) \setminus \bigcup_{j<m} \setD_{c_j}$ into
$\setD_{c_m}$.
\end{enumerate}
For the step 2 (a), a simple way is to set each 
$b_i = |\setS \setminus \setC|^\gamma / m$, where $0 < \gamma \leq 1$. 
Setting $\gamma = 1$ gives a balanced tree.
However, as is shown by the experimental results
(Sec.~\ref{sec:exp}), unbalanced ball partitioned tree can be much more efficient in 
some search tasks, especially in high dimensions and / or large query radii.
The reason for this is that in high dimensions the distance distributions are very concentrated,
and unbalancing effectively makes the covering balls smaller, and hence the probability of 
intersecting with each other and the query ball decreases. 
Another reason is that as the right-most path of the tree gets longer, the clusters get 
more compact and more distances are evaluated in the preprocessing phase, which pays off 
at the search phase.

Note that when using constant arity $m$ in all nodes, and $\gamma < 1$ does not affect the 
space complexity, but increases the preprocessing time. For non-constant arities of the form
$n_i^\alpha$ both space and preprocessing costs are increased quite significantly. We do not give 
the formulas here, but just refer to the experimental results in Sec.~\ref{sec:exp}.

It is also interesting to notice that using GNAT with ball partitioning, huge arity for the root 
and/or large buckets for the leaves we obtain a tree of height 1, which then resembles
List of Clusters (LC) \cite{lc} with range tables added on top (i.e.\ LC uses 
only $R^{hi}_{i,i}$ values for pruning). Another way to obtain a similar data structure (plus the 
added range tables), is to use small to moderate constant arity $m$ and quite small $\gamma$, 
and use bucketing for subsets of size $|\setS \setminus \setC|^\gamma / m$.

We remark that methods like excluded middle point (ball) partitioning \cite{vpf} 
could also be adapted to work with GNAT with some effort, but we leave that as a future work.

\subsection{Reducing space} \label{sec:smallspace}

The space complexity can be controlled by adjusting the arity, either using the fixed arity
$m$ or using the dynamic method as described in Sec.~\ref{sec:powa}. However, this also
affects the search cost at the same time, possibly by a large factor. 
Moreover, in external memory EGNAT the arity $m$ is limited by the disk block size.
We show how fixed-point representation can be used to reduce the space, as well as 
give another method that enables controlling the arity and the size of the range tables
somewhat independently.

\subsubsection{Fixed-point representation} \label{sec:fp}

We suggest an implementation method that can compress the distance range matrix without 
reducing its dimensions (i.e.\ tree arity).

The distance range entries in $R$ are either continuous or discrete, depending on the metric space,
and typically stored using e.g.\ {\tt float}s (4 bytes) or {\tt int}s (typically 4 bytes) 
in C/C++/Java-programming languages.
If in the discrete case the distance range is small (such as edit distance of relatively short strings),
then using smaller integer type (such as {\tt char}) may suffice, or one can use custom coded smaller
integer representations (e.g.\ use 4 high-order bits of the $lo$ and $hi$ values and encode them into a 
single 8-bit byte).

In case of continuous distance values, one can resort to fixed-point representation. That is, store
the $R_{i,j}$ entries in fixed-point representation in the preprocessing step, and when the values
are needed in the search phase, retrieve and convert back to floating point.
Note that we need not to do any arithmetic in the fixed-point representation.

We use the notation \fxp{m}{b} to denote a fixed-point type that is encoded with $b$ bits, $m$ bits
are reserved for the magnitude (unsigned integer) part, and $b-m$ for the fractional part. 
For example, \fxp{2}{8} stores the number in 8-bits (one byte), where 2 bits are reserved for the 
integer part and 6 bits for the fractional part. This could be used to represent the distances in 
10 dimensional unitary cube with Euclidean distance.

Conversions are easy. 
If $x$ is a floating point type and $y$ a fixed point type, then we can do the
conversions as $y \leftarrow x 2^{b-m}$ and $x \leftarrow y / 2^{b-m}$, where we assume truncation
as the rounding mode for the least significant fraction bit when converting to integer type. 
The value $2^{b-m}$ is called a scaling factor.
If we want to always round up the least significant fraction bit, then we can simply 
do $y \leftarrow x 2^{b-m}+1$. Indeed, in order to the GNAT search algorithm to work correctly, 
we need to round up the $hi$ values. Given $[lo,hi]$ in floating point representation, 
we actually store
\[
R_{i,j} \leftarrow [lo\; 2^{b-m}, hi\; 2^{b-m} + 1].
\]
That is, round $lo$ down and $hi$ up.

One problem with fixed-point representation is that we cannot have large magnitude and good precision
with a small number of bits, which is a problem if the distance values can be sometimes large and 
sometimes small. The other problem is more implementation specific, i.e.\ how to fix \fxp{m}{b}
(this could be done dynamically, however). One solution that works quite well for a lot of different
scenarios is to use some kind of range transform. For example, one could convert $\log x$ into 
fixed-point representation instead of converting plain $x$.
Again, notice that this is not a problem as we do no arithmetic in fixed-point representation.
However, $\log x$ is suitable only if $x>1$. Better method is to use $x^\beta$, for 
some $0 < \beta < 1$, as this transforms all positive numbers towards $1$. 
On the other hand, using very small $\beta$ would mean too much loss of precision. 
In practice values like $\beta=1/5$ work very well for \fxp{2}{8}. 
The conversion becomes now
\[
R_{i,j} \leftarrow [lo^\beta\; 2^{b-m}, hi^\beta\; 2^{b-m} + 1].
\]
To convert a fixed point value $y$ back to floating point we do 
$x \leftarrow (y / 2^{b-m})^{1/\beta}$

As shown in the experimental results, using just one byte to store the (continuous) distance values
gives negligible performance loss while reducing the space by a factor of 4. In some 
cases using fixed-point instead of floating point actually increases the performance 
(i.e.\ CPU time)
a little, probably 
due to better cache utilization.

\subsubsection{Smaller range tables} \label{sec:smalltables}

Another idea is to have smaller range tables by not (fully) indexing every center. 

Note that EGNAT uses just one column of $R$ (corresponding to the nearest neighbor of $q$ in $\setC$) 
in each node during searching. 
This gives the idea of limiting the set of centers where the nearest neighbor can be 
selected, effectively removing some of the columns from $R$.
That is, we can select a subset $\setCp \subseteq \setC$, and compute $|\setC| \times |\setCp|$ sized 
distance range tables. This does not affect the arity of the tree, just the pruning process, which is
trivial to adapt in the case of EGNAT. For GNAT we can replace the AESA-like algorithm e.g.\ with a
LAESA-like algorithm \cite{laesa}. 
Preprocessing time for the $R$ tables is also improved.

In any case, this can make the search algorithm potentially worse, i.e.\ it may not prune the tree 
as effectively now, but in return the tree arity can be larger thanks to the smaller tables. 
The arities can be increased by a factor $\sqrt{a}$, for $a>1$, if $|\setCp| = |\setC| / a$, while 
keeping the same memory usage for the tables (per node). 
This again makes the clusters smaller, giving an opportunity to more
effective pruning.
This method may have a positive effect especially in secondary memory implementation.

\section{Experimental results} \label{sec:exp}

We have implemented the algorithms in C and ran various experiments with different data sets.
We used random vectors in uniformly distributed unitary cube as well as 112 dimensional
color histograms, both with Euclidean distance, as well as an English dictionary and a larger 
dictionary (combined from several languages, duplicates removed) with edit-distance.
The databases are from \cite{LibMetricSpace}.
In each case we picked 1000 objects randomly and used them as the queries, building the database
using the rest. In each case the index is built the whole way down, i.e.\ no bucketing was used 
for the leaves.
Pivots were selected in random in all cases. 
We call our algorithm GNATTY in what follows. 

\begin{figure}[!ht]
\centerline{\includegraphics[scale=0.95]{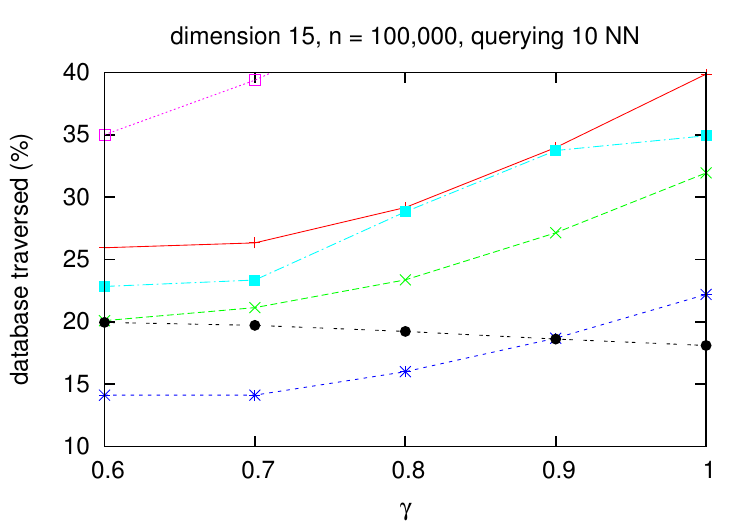}}
\centerline{\includegraphics[scale=0.95]{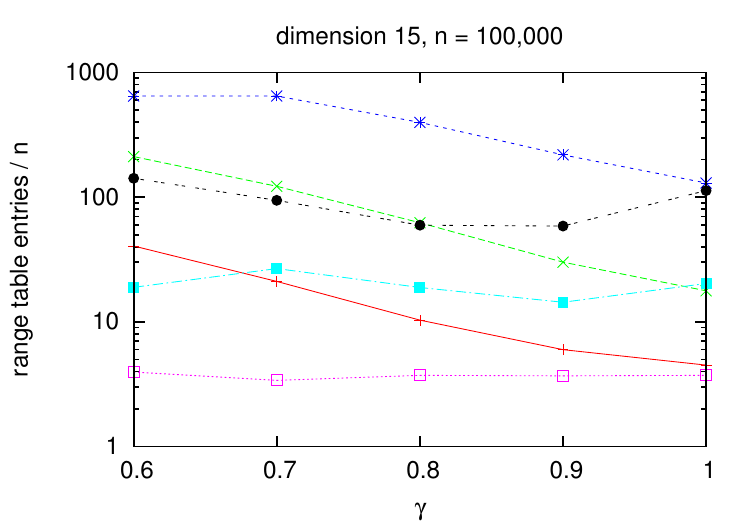}}
\centerline{\includegraphics[scale=0.95]{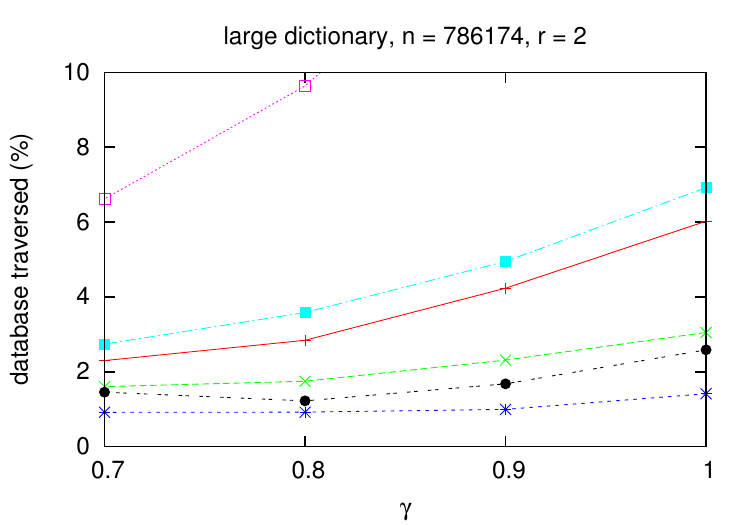}}
\centerline{\includegraphics[scale=0.95]{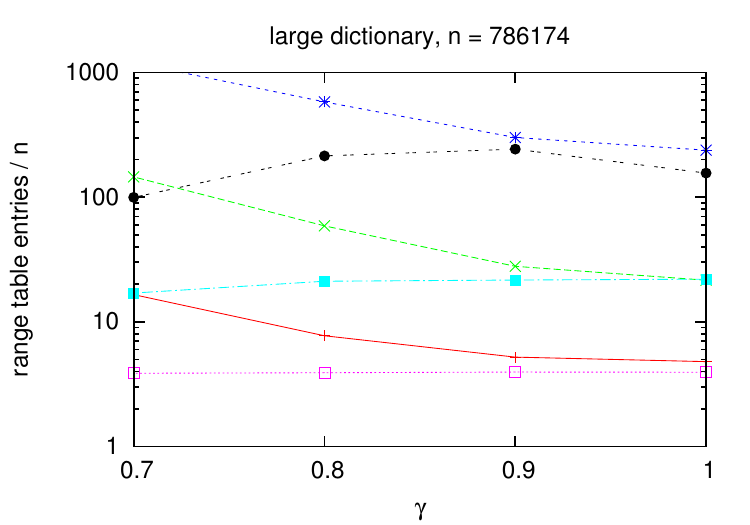}}
\centerline{\includegraphics[scale=0.95]{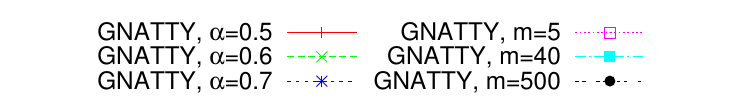}}
\caption{
Unbalanced ball partitioning, using fixed and variable arities. 
1st plot: 
distance evaluations for random vectors for different $\gamma$, range query retrieves
10 neighbors;
2nd plot:
the number of range tables entries corresponding to the previous plot;
3rd and 4th plots:
as above, but for string dictionary and $r = 2$.
}
\label{fig:gamma}
\end{figure}

We used both hyperplane partitioning (as in original GNAT) and (unbalanced) ball partitioning. 
For ball-partitioning, the optimal value of $\gamma$ (see Sec.~\ref{sec:ballp}) depends on the 
dimensionality of the space and the selectiveness of the queries, as well as the arities. 
In particular, for the ``easy''
cases the optimum is $\gamma = 1$, and it decreases as the queries become ``harder''.
Fig.~\ref{fig:gamma} shows two cases (random vectors in 15 dimensional space and a string 
dictionary) where it is beneficial to use $\gamma < 1$. Note that the space complexity is 
also affected, in particular for non-constant arities and large $\alpha$, which means that 
in most cases the optimum $\gamma$ may be impractical. In general, keeping $\gamma$ close to $1$
and adjusting $\alpha$ gives better control for the space/time trade-offs.
In all the subsequent plots we use 
a fixed $\gamma = 0.9$, as it usually gives quite a noticable 
performance boost, while not affecting the space 
complexity when using non-constant arities too much.

\begin{figure}[!ht]
\centerline{\includegraphics[scale=0.95]{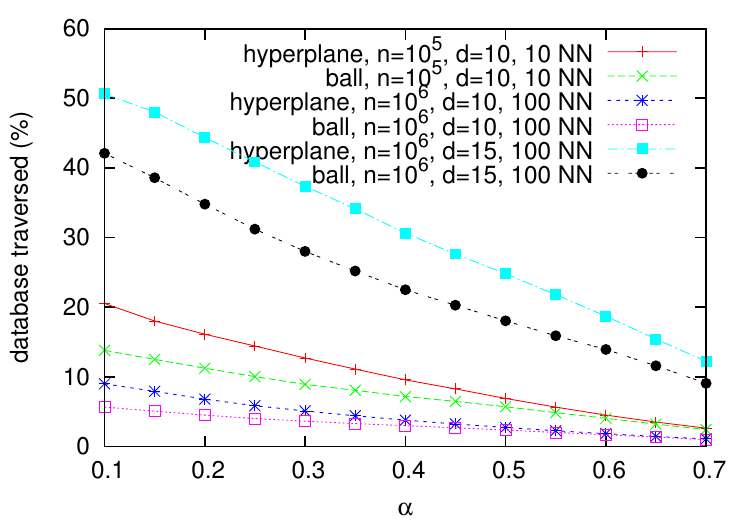}}
\centerline{\includegraphics[scale=0.95]{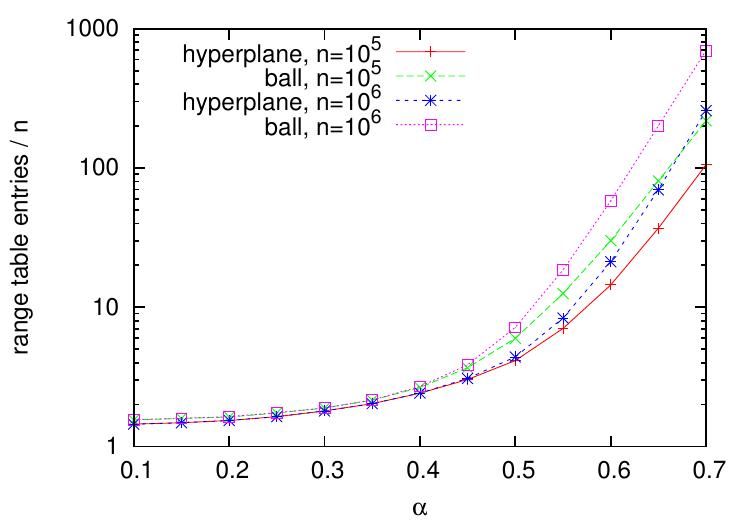}}
\centerline{\includegraphics[scale=0.95]{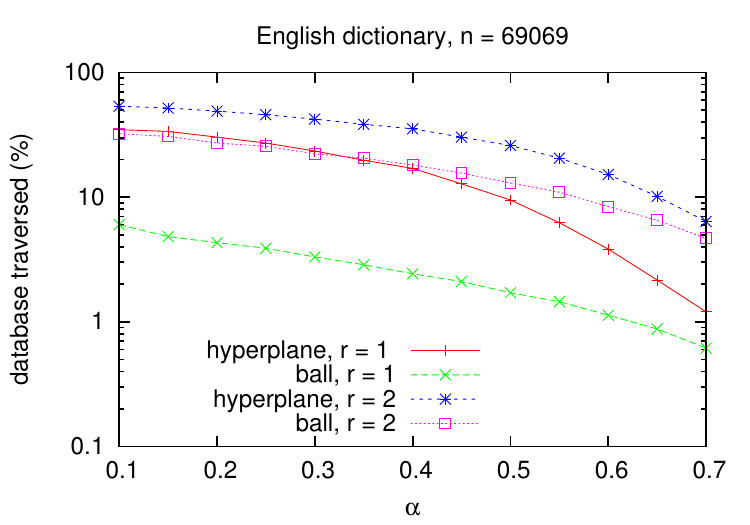}}
\centerline{\includegraphics[scale=0.95]{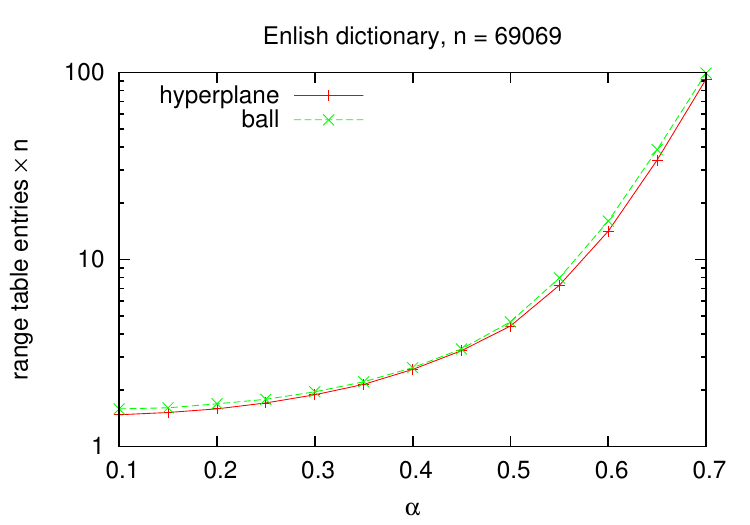}}
\caption{
1st plot: 
distance evaluations for random vectors for different $\alpha$, range query retrieves
10 or 100 nearest neighbors;
2nd plot:
the number of range tables entries corresponding to the previous plot;
3rd and 4th plots: as the previous two, but for strings.
}
\label{fig:alpha}
\end{figure}

Fig.~\ref{fig:alpha} shows the effect of $\alpha$ 
for two synthetic vector spaces and for English dictionary. The space is very close to 
$O(n \log\log n)$ for $\alpha = 0.5$, but starts to increase rapidly after that.
Note however that the data itself can take a lot of space; 
e.g.\ vectors in 15 dimensional space (using one {\tt float} per 
coordinate) requires $60n$ bytes, which is easily more than what the range tables require for 
moderate $\alpha$. In any case, if there are available memory, increasing $\alpha$ reduces
the number of distance evaluations steadily. 
Observe that ball partitioning gives better results than the original hyperplane partitioning,
especially for strings.

\begin{figure}[!ht]
\centerline{\includegraphics[scale=0.95]{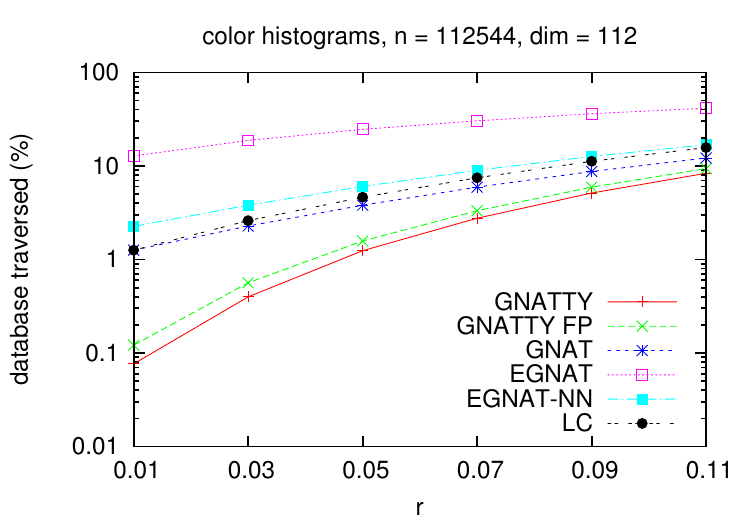}}
\centerline{\includegraphics[scale=0.95]{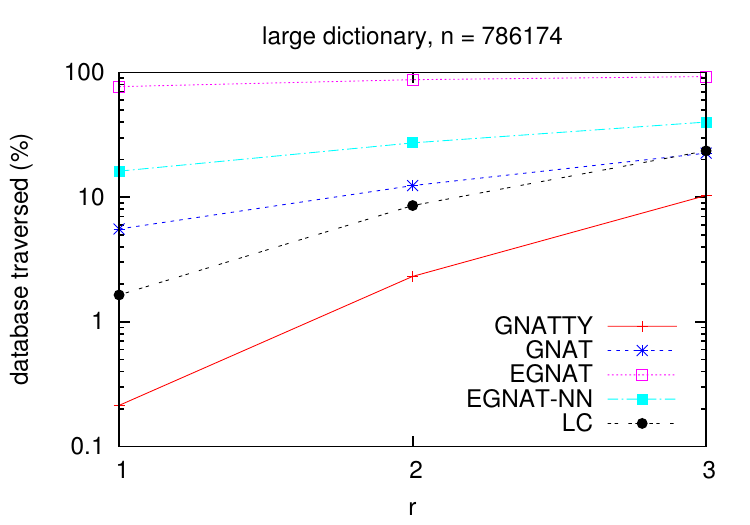}}
\caption{Top: GNATTY and GNATTY FP (Fixed-Point), both with ball partitioning, use $\alpha = 0.5$, 
the other GNAT variant use constant arity that results
in the same memory consumption as in GNATTY, except  
GNATTY FP uses 1/4th of the memory GNATTY uses.
LC uses linear space.
Bottom: as above, but GNATTY uses $\alpha=0.6$.
}
\label{fig:eqmem}
\end{figure}

Fig.~\ref{fig:eqmem} compares GNATTY (using ball partitioning) against the original GNAT 
(hyperplane partitioning and constant arity) and two variants of EGNAT, so that all methods 
use the same amount of memory. We also compare against GNATTY that uses fixed-point (FP) 
(see Sec.~\ref{sec:fp})
to store
the range tables (1 byte per distance; the baseline method uses 1 {\tt float}, i.e.\ 4 bytes).
Recall that EGNAT (besides the added dynamism and external memory implementation) is as GNAT 
with simpler pruning rules. As seen in the plots, this does not work well for large arities
(the original EGNAT uses relatively low arities). Hence we added a nearest neighbor (NN) index 
over the pivots so that the nearest pivot (along with any pivot in the range) 
to the query can be retrieved faster. 
The performance of GNATTY FP is close to GNATTY, even if the former uses only 1/4th of the space 
and approximated distance values.
We include List of Clusters (LC) \cite{lc} as a baseline competitor. LC uses only $O(n)$ space.
The bucket size for LC was optimzed for $r = 0.05$ and $r = 2$, for color histograms and strings,
respectively.
 
As an other example, using $r = 0.03$ on the color histograms database, GNAT with hyperplane 
partitioning would need $m \approx 135$ to reach the performance of GNATTY with ball partitioning
and $\alpha = 0.5$. On the large string dictionary for $r = 2$, 
GNAT would need $m \approx 1120$ to match GNATTY with $\alpha = 0.6$. Note that the constant factor 
in the $(nm)$ space complexity is often relatively small, as near the leaves
it is not possible to use the full arity as there are not enough objects left. E.g., for 
$m=1120$ and the strings dictionary, GNAT requires ``only'' about $600n$ range table entries.

We also ran preliminary experiments on using smaller range tables (see Sec.~\ref{sec:smalltables}). 
As expected, this reduces the performance, some of which can be bought 
back by using larger arities (sometimes the performance is improved a bit). 
The net effect is that using the same space the tree
height can be reduced, but the queries become somewhat slower, and this effect increases 
the smaller the range tables become. We omit the plots. Nevertheless, the technique has some 
promise for external memory implementation, which is a subject of future work.

\section{Concluding remarks}

We have shown several methods how to improve GNAT and verfied their practical performance. 
However, there are many possibilities for further work. 
\begin{itemize}
\item 
The hyperplane partitioning construction cost can be lowered somewhat by using an auxiliary 
index to solve the 1-NN queries in step 2 of the construction algorithm, especially for 
high arities. That is, build 1-NN index for the centers / pivots, and use 1-NN queries for each
object to find its associated center.
\item 
Bulk loading the tree can also be lazy, i.e.\ a branch of the tree can be built
only on demand, when the search algorithm enters it, which amortizes the search and construction
costs. 
\item
The range tables for the nodes can be also built in the same spirit as the previous item, 
i.e.\ any 
$R_{i,j}$ value can be initialized to some default value and the real value is computed when 
it is needed the first time. This can be also used with the EGNAT insertion 
algorithm to amortize its cost; i.e.\ 
new elements are inserted into leaves, which are initially buckets and promoted to 
full GNAT like internal nodes when they becomes full.
\item
GNATTY techniques can be used for external memory implementation as well.
EGNAT uses the same arity for all internal nodes (including root), depending on the disk block 
size. However, the root node can be made (much) larger than the other nodes, as it can be 
kept in main memory all the time. 
\end{itemize}


\end{document}